\documentclass[a4paper]{article}
\usepackage[utf8]{inputenc}

\usepackage{amsfonts,amsmath,amssymb}
\usepackage{booktabs} 
\usepackage[lined,boxed,noend,ruled,resetcount]{algorithm2e}

\usepackage{graphicx}
\usepackage{caption}
\usepackage{subcaption}
\usepackage{setspace}
\usepackage{marginnote}
\usepackage{changes}
\usepackage{url}
\usepackage{a4wide}
\usepackage{times}

\begin{document}
\title{Accelerating Energy Games Solvers\\ on Modern Architectures\thanks{This research is supported by INdAM-GNCS grants
and by \textsc{YASMIN}~(RdB-UniPG2016/17) project}}

\author{Andrea Formisano\\
{Universit{\`a} di Perugia}\\
\url{andrea.formisano@unipg.it}
\and Raffaella Gentilini\\
  {Universit{\`a} di Perugia}\\
\url{raffaella.gentilini@unipg.it}
\and Flavio Vella\\
  {dividiti Ltd}\\
\url{flavio@dividiti.com}}

\date{}

\newtheorem{lemma}{Lemma}
\newtheorem{theorem}[lemma]{Theorem}
\newtheorem{definition}[lemma]{Definition}
\newtheorem{example}[lemma]{Example}

\newcommand{\nat}{\mathbb N}
\newcommand{\zed}{{\mathbb Z}}
\newcommand{\rat}{{\mathbb Q}}
\newcommand{\posrat}{{\mathbb Q}^{\geq 0}}
\newcommand{\sposrat}{{\mathbb Q}^{> 0}}
\newcommand{\real}{{\mathbb R}}
\newcommand{\posreal}{{\mathbb R}^{\geq 0}}
\newcommand{\sposreal}{{\mathbb R}^{> 0}}
\newcommand{\M}{\mathcal{M}}
\newcommand{\all}{\stackrel{\forall}{\rightarrow}}
\newcommand{\some}{\stackrel{\exists}{\rightarrow}}
\newcommand{\mathR}{\mathbb{R}}
\newcommand{\mathN}{\mathbb{N}}
\newcommand{\CTL}{{\sf CTL}}
\newcommand{\ce}{\stackrel{\delta}{\rightarrow}}
\newcommand{\te}{\stackrel{t}{\rightarrow}}
\newcommand{\ca}{\stackrel{a}{\rightarrow}}
\newcommand{\de}{\stackrel{e}{\rightarrow}}
\newcommand{\Inv}{\textit{\mbox{Inv}}}
\newcommand{\Init}{\textit{\mbox{Init}}}
\newcommand{\NF}{\textit{\mbox{NF}}}
\newcommand{\must}{\tiny{\textit{\mbox{must}}}}
\newcommand{\game}{$\Gamma=(V, E, w, \langle V_0,V_1\rangle)$}
\newcommand{\pre}{{\sf pre}}
\newcommand{\post}{{\sf post}}
\newcommand{\val}{{\sf val}}
\newcommand{\credit}{{\sf c}}
\newcommand{\MPG}{{\sf MPG}}
\newcommand{\EG}{{\sf EG  }}
\newcommand{\Path}{{\sf AcyclicPath}}
\newcommand{\counter}{{\sf count}}
\newcommand{\abs}[1]{\ensuremath{\lvert #1\rvert}}
\newcommand{\tuple}[1]{\langle #1 \rangle}
\newcommand{\outcome}{\mathsf{outcome}}
\newcommand{\MeanPayoff}{\mathsf{MP}}
\newcommand{\LimAvg}{\mathsf{LimAvg}}



\maketitle

\begin{abstract}
Quantitative games, where quantitative objectives are defined on weighted game arenas, provide natural tools for designing faithful models of embedded controllers. Instances of these games that recently gained interest are the so called Energy Games.  
The fast-known algorithm solves Energy Games in $\mathcal{O}(EVW)$ where $W$ is the maximum weight. Starting from a sequential baseline implementation, we investigate the use of massively data computation capabilities supported by modern Graphics Processing Units to solve the \emph{initial credit problem} for Energy Games. 
We present four different parallel implementations on multi-core CPU and GPU systems. Our solution outperforms the baseline implementation by up to 36x speedup and obtains a faster convergence time on real-world graphs.
\end{abstract}

\section{Introduction}\label{sec:introduction}

  Classic game theoretic-formulations of the control synthesis problem rely on two player  zero-sum  games on   graphs, where  the system is opposed to an antagonist environment.
  In this context,  the modeling game arena is a  graph, where  the vertices are either owned by  player~$0$ (the system) or by the antagonist player~$1$ (the environment).
  The two players move a pebble along the vertices of the graph, starting from an initial position.
  Whenever the pebble is on a vertex belonging to player~$0$ (resp. player~$1$), the latter decides where to move the pebble next, according to his strategy.
  The infinite path followed by the pebble is called a play and represents one possible behavior of the system.
  The winning objective for player~$0$ (a set of plays)  encodes exactly the acceptable behaviors of the system.
  Therefore, the goal of player~$0$ is to ensure with his strategy---the synthesized controller---that the outcome of the game is an acceptable behavior of the system, whatever the strategy played by his adversary.

  Quantitative games, where quantitative objectives are defined on \emph{weighted} game arenas,  provide natural tools for designing faithful models of embedded controllers, since they allow to explicitly handle the quantitative constraints imposed by the environment, the lack of resources or the targeted parameters of operability.
  In the late seventies, traditional game theory developed for economics defined a number of nowadays classic quantitative objectives, such as meanpayoff (MPG)  and discounted-payoff~\cite{Aptbook,ZP}, that have been recently extensively investigated for the specification and design of reactive systems~\cite{Aptbook}.
  In turn, the problem of controller synthesis with resource constraints has inspired  new quantitative objectives and   quantitative games, such as e.g.~so called energy games in \cite{CAS03,BFLMS08,BrimEtAl} and their variants (see e.g. \cite{CD10,CDHR10,FJLS11,BMRLL15,VetAl15}). The latter  turn out  to be of broad interest, having applications in computer aided synthesis \cite{CD10,CDHR10,Bruyere17}, real-time systems 	\cite{BCR14,BMRLL15}, as well as economy, due to their connection with meanpayoff and discounted-sum objectives~\cite{Aptbook}.

  In \emph{energy games},  edges are fitted with integer weights aimed at modeling  rewards or costs.
  The objective of player~$0$ is to maintain the sum of the weights (called the energy level) always  positive along the play, given a fixed initial credit of energy.
  Energy games were introduced in \cite{CAS03,BFLMS08}, where they were also proven  memoryless determined:  namely, each vertex is either  winning for player~$0$ or it is winning for player~$1$, and memoryless strategies are sufficient to consider.
  Deciding whether a vertex $v$ is winning for player~$0$  in an energy game is equivalent to the corresponding problem in meanpayoff games, and the latter equivalence has provided faster pseudo-polynomial algorithms for  MPGs \cite{BrimEtAl,CHN14,CR17}.
  The above decision problems lie notoriously  in the complexity class NP $\cap$ coNP (and even UP  $\cap$ coUP), while finding polynomial time procedures for them  is a long standing open problem \cite{ZP,Aptbook}.
  The minimum credit problem on energy game subsumes the corresponding decision problem,  and  asks the following:  to determine, for each vertex $v$ of an energy game $\mathcal{G}$, whether $v$  is winning for player~$0$ and which is the minimum credit to stay alive along each play starting from $v$.
  Such a problem can be also solved in pseudo-polynomial time \cite{ZP,BrimEtAl}.
  Recently, parallel architectures like Graphics Processing Units (GPU) have been successfully used in accelerating many irregulars and low-arithmetic intensity applications like graph traversal-based algorithms, in which the control flow and memory access patterns are data-dependent \cite{burtscher2012quantitative, Bernaschi2016}. 
  Motivated by the large instances that naturally arise from the specification,  design and control  of reactive systems, in this work we investigate the use of massively data computation capabilities supported by modern GPUs for solving the initial credit problem on energy games. Also, to alleviate the workload unbalancing among threads, we propose a suitable data-thread mapping technique which allows to efficiently solve traditional Energy Games instances.
  The contributions of the paper are manifold:
  \begin{itemize}
  \item
  we provide a parallel implementation, exploiting traditional multi-core architectures, of the state-of-the-art initial credit procedure for enrgy games in \cite{BrimEtAl};
  \item
  we developed a CUDA implementation which relies on a traditional vertex-parallelism approach and a more suitable variant based on warp-centric parallelism;
  \item
  we report experimental results where we compare the performance achieved by the above mentioned implementations and a completely sequential one.
  \end{itemize}
  After reviewing some minimal preliminary notions (Section~\ref{sec:preliminaries}),
  the theoretical results on Energy Games (EG) relevant for this paper are recalled in Section~\ref{sec:games}.
  Sections~\ref{sec:multicore}-\ref{sec:implementazione} describe our parallel solutions.
  The results of the experimentation activity and a comparison between the 
  solvers are outlined in Section~\ref{sec:esperimenti}. We finally discuss related works in Section~\ref{related} and draw our conclusion in Section~\ref{conclusions}.

\section{Preliminaries}\label{sec:preliminaries}

\subsubsection*{Weighted graphs}
A \emph{weighted graph} is a tuple $G=(V, E, w)$, where  $V$ is a set of of vertices,
 $E \subseteq V \times V$ is a set  of edges, and  $w: E \to \zed$ is a weight function
assigning an integer weight to each edge.  
We assume that weighted graphs are \emph{total}, i.e. for all $v \in V$,
there exists $v' \in V$ such that $(v,v') \in E$.
Given a set of vertices $U \subseteq V$ in a weighted graph $G=(V, E, w)$, we denote by $\pre(U)$ the set of vertices
having a successor in $U$, i.e. $\pre(U)= \{v \mid \exists v' \in U: (v,v') \in E \}$,
and by $\post(U)$ the set of successors of vertices in $U$, i.e.
$\post(U)=\{v \mid \exists v' \in U: (v',v) \in E \}$.

A (finite) path $p$ in $G=(V, E, w)$ is a nonempty sequence of vertices $v_0 v_1 \dots $ (resp. $v_0 v_1 \dots v_n$) such that
$(v_i,v_{i+1}) \in E$ for all $0 \leq i$ (resp. $0 \leq i < n$).
The length of a finite path  $p=v_0 v_1 \dots v_n$ is the number of vertices $n+1$ in $p$, denoted $|p|$.
Given a (finite) path $p$ and an integer $j\geq 0$ (resp. $0\leq j\leq n$), we denote by $p^j$ the prefix $v_0\dots v_j$ of $p$ up to $j$ and by $p[j]$ the vertex $v_j$.
A \emph{cycle} in $G=(V, E, w)$ is a finite path $p=v_0 v_1 \dots v_n$ such that $n \geq 1$ and $v_0=v_n$.
We say that a cycle in a weighted graph is \emph{negative} (resp. \emph{nonnegative})
if the sum of its edge weights is less than $0$ (resp. not less than $0$).
Given $v\in V$, a cycle $v_0 v_1 \dots v_n$ is said \emph{reachable} from $v$ in $G$ if there exists a path $u_0 u_1 \dots u_m$ in $G$ such that $u_0=v$ and $u_m=v_0$.
A path $v_0 v_1 \dots v_n$ is \emph{acyclic} if $v_i \neq v_j$ for all $0 \leq i < j \leq n$.

\subsubsection*{Graph games}
A \emph{game arena} is a tuple $\Gamma = (V, E, w, \tuple{V_0, V_1})$
where $G^\Gamma=(V, E, w)$ is a weighted graph and $\tuple{V_0, V_1}$ is a partition
of $V$ into the set $V_0$ of player-$0$ vertices and the set $V_1$ of player-$1$ vertices.
An \emph{infinite game} on $\Gamma$ is played for infinitely many rounds
by two players moving a pebble along the edges of the game arena $G^\Gamma$.
In the first round, the pebble is on some vertex $v \in V$.
In each round, if the pebble is on a vertex $v\in V_i$ ($i=0,1$), then player~$i$ chooses
an edge $(v,v') \in E$ and the next round starts with the pebble on $v'$.
A \emph{play}  in the arena $\Gamma = (V, E, w, \tuple{V_0, V_1})$ is an infinite 
path $\pi\subseteq V^\omega$ in $(V,E,w)$.
An \emph{objective}  for player $0$ is a set $\mathcal{O}_0\subseteq V^\omega$: the play $\pi\in V^\omega$ is said to be winning for player $0$ if $\pi\in\mathcal{O}_0$.
In this paper, we restrict our attention to $0$-sum games, i.e. games where the two players are antagonists: therefore, the objective of player $1$ is $\mathcal{O}_1=V^\omega\setminus \mathcal{O}_0$.
A \emph{game}  is a tuple $\mathcal{G}=\langle \Gamma, \mathcal{O}_0\rangle$, where $\Gamma$ is a graph arena, and $\mathcal{O}_0\subseteq V^\omega$ is the \emph{objective} of player $0$.
Given a game $\langle \Gamma, \mathcal{O}_0\rangle$, the players play according to strategies to ensure a play that accomplish their objective.

A \emph{strategy} for player~$i$ ($i=0,1$) is a function $\sigma: V^* \cdot V_i \to V$,
such that for all finite paths $v_0 v_1 \dots v_n$ with $v_n \in V_i$,
we have $(v_n,\sigma(v_0 v_1 \dots v_n)) \in E$.
We denote by $\Sigma_i$ ($i=0,1$) the set of strategies for player~$i$.
A strategy $\sigma$ for player~$i$ is {\em memoryless}  if $\sigma(p)=\sigma(p')$
for all sequences $p=v_0 v_1 \dots v_n$ and $p' = v'_0 v'_1 \dots v'_m$ such that $v_n=v'_m$.
We denote by $\Sigma^M_i$ the set of memoryless strategies of player~$i$.
A play $v_0 v_1 \dots v_n \dots$ is \emph{consistent} with a strategy $\sigma$ for player~$i$
if $v_{j+1} = \sigma(v_0 v_1 \dots v_j)$ for all positions $j \geq 0$ such that $v_j \in V_i$.
Given an initial vertex $v \in V$, the \emph{outcome} of two strategies $\sigma_1 \in \Sigma_1$ and $\sigma_2 \in \Sigma_2$ in $v$
is the (unique) play $\outcome^\Gamma(v,\sigma_0,\sigma_1)$ that starts in $v$ and
is consistent with both $\sigma_0$ and $\sigma_1$.
Given a memoryless strategy~$\pi_i$ for player~$i$ in the game $\mathcal{G}=\langle \Gamma, \mathcal{O}_0\rangle$,
we denote by $G^\Gamma(\pi_i) = (V, E_{\pi_i}, w)$ the weighted graph
obtained by removing from $G^\Gamma$ all edges $(v,v')$ such that
$v \in V_i$ and $v' \neq \pi_i(v)$.

\section{Energy Games}\label{sec:games}
In this section, we  introduce  energy games  \cite{CAS03,BFLMS08,BrimEtAl},  that are the  main objective of the rest of this paper.
An energy game is a  game   $\mathcal{G}=\langle \Gamma,\mathcal{O}_0\rangle$ over the arena $\Gamma = (V, E, w, \tuple{V_0, V_1})$, where the goal of player~$0$ is to
construct an infinite play $v_0 v_1 \dots v_n \dots$ such that for some \emph{initial credit} $c \in \nat$, it holds that:
\begin{equation}\label{wincon}
c + \sum_{i=0}^{j} w(v_i,v_{i+1}) \geq 0 \text{ for all } j \geq 0
\end{equation}
\noindent
The quantity $c + \sum_{i=0}^{j-1} w(v_i,v_{i+1})$  in (\ref{wincon}) is called the
\emph{energy level} of the play prefix $v_0 v_1 \dots v_{j}$, given  the \emph{initial credit} $c$.
Conversely, player $1$ aims at building a
play $\pi=v_0 v_1 \dots v_n \dots$ such that for any initial credit $c$, there exists a prefix
$\pi^j$ of $\pi$ such that the energy level of $\pi^j$ is negative.
Formally, energy games are defined as follows: 


\begin{definition}[Energy Games]\label{defEG}
An energy game ({\sf EG}) is a  game 
$\mathcal{G}=\langle \Gamma,\mathcal{O}_0\rangle$, where $\Gamma = (V, E, w, \tuple{V_0, V_1})$ and $\mathcal{O}_0$ is given by:
$$\begin{array}{l}\mathcal{O}_0\,=\{\pi|\: \pi \mbox{  is a path in $G^\Gamma=\langle V,E,w\rangle$} \,\wedge  
\\
\phantom{\mathcal{O}_0\,=\{\pi|\: }\exists c\in \nat  : c + \sum_{i=0}^{j-1} w(v_i,v_{i+1}) \geq 0 \text{ for all } j \geq 0 \}
\end{array}$$
\end{definition}
A vertex $v \in V$ is \emph{winning} for player~$i$, $i\in\{0,1\}$ if there exists an initial
credit $c(v)$ and a winning strategy for player~$i$ from $v$ for credit $c(v)$.
In the sequel, we denote by $W_i$ the set of winning vertices for player $i$.

\begin{figure}[tb!]
{\centering
\includegraphics[scale=0.5]{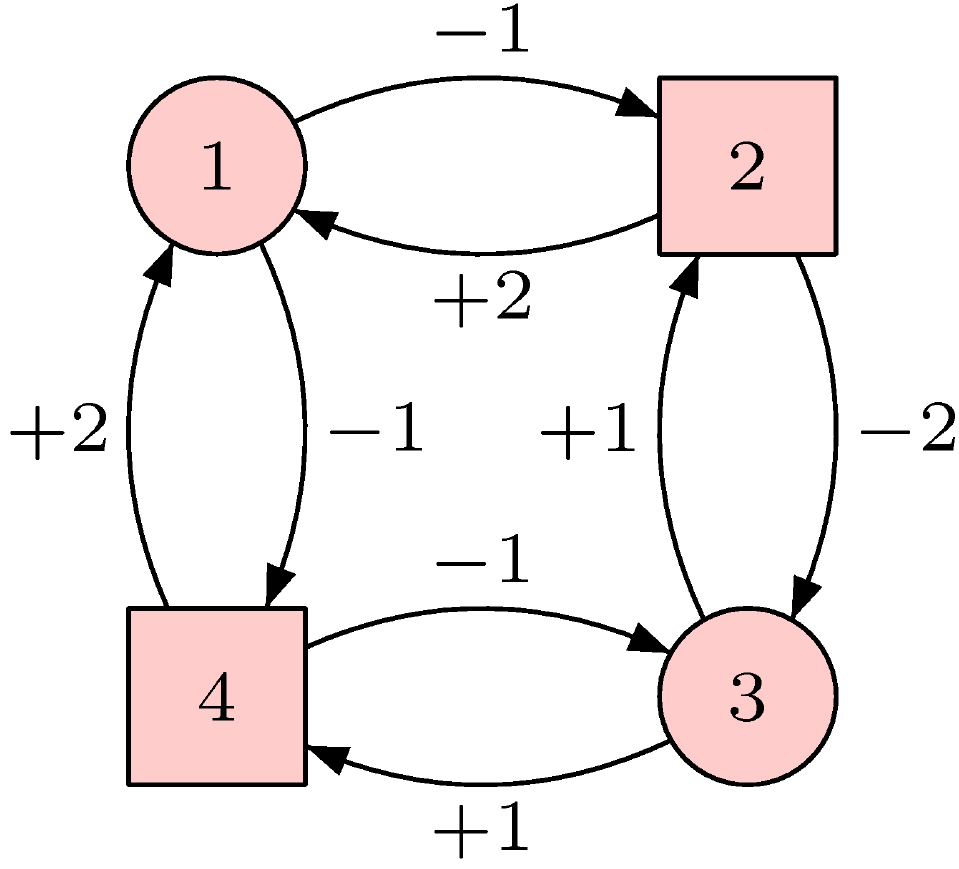}
\caption{\label{fig1}An Energy Game $\mathcal{G}$.}
}
\end{figure}

Energy games are memoryless determined~\cite{BFLMS08},
i.e. for all $v\in V$, either $v$ is winning for player~$0$,
or $v$ is winning for player~$1$, and memoryless strategies are sufficient.
Using the memoryless determinacy of energy games, one can easily prove
the following result for {\sf EG}, characterizing the winning strategies for player $0$ in a {\sf EG}.
\begin{lemma}[\cite{BrimEtAl}]\label{EGcycles}
Let $\mathcal{G}=\langle \Gamma,\mathcal{O}_0\rangle$ be an {\sf EG}, for all vertices $v \in V$,
for all memoryless strategies $\pi_0 \in \Sigma^M_0$ for player~$0$, the strategy $\pi_0$ is winning from $v$
iff all cycles reachable from~$v$ in the weighted graph $G^\Gamma(\pi_0)$ are nonnegative.
\end{lemma}
\begin{example}
Consider the energy game depicted in Figure~\ref{fig1}, where round vertices are owned by player $0$, and square vertices are owned by player $1$. 
The set of winning vertices for player $0$ is $W_0=V$.
In fact,  the memoryless strategy $\sigma_0: V_0\rightarrow V$ for player $0$, where $\sigma_0(1)=\sigma_0(3)=4$, ensures that any cycle in a play consistent with $\sigma_0$---against any (memoryless) strategy $\sigma_1$ for player $1$---is nonnegative.
Therefore, provided a suitable initial  credit of energy, each play consistent with  $\sigma_0$  will enjoy a nonnegative energy level along its run.
Given $v\in W_0=V$, the minimum initial credit $c(v)$   that player $0$ needs to survive along a play starting from $v$  is given by $c:W_0\rightarrow \mathbb{N}$, where  $c(1)=2, c(2)=2, c(3)=1, c(4)=1$.
As a further example, if the edge $(3,4)$ is deleted from the energy game in Figure~\ref{fig1}, then player $0$ does not have any winning strategy from $W_1=\{2,3\}$, but only from the set of vertices $W_0=\{1,4\}$  with initial credit of energy $c(1)=1, c(4)=0$.
\end{example}

The next definition introduces the \emph{initial credit} problem. 
\begin{definition}[Initial Credit Problem]\label{defInitCreditProblem}
Given an energy game $\mathcal{G}=\langle \Gamma,\mathcal{O}_0\rangle$, the  \emph{initial credit problem} on $\mathcal{G}$  asks to determine,   for each vertex $v\in V$, the following:
\begin{enumerate}
\item  if $v$ is winning for player~$0$, i.e. if $v\in W_0$.
\item  in case $v\in W_0$, 
the minimum initial credit $\credit(v)$ such that there is a winning strategy $\sigma_0$ for player~$0$ in~$\mathcal{G}$.
\end{enumerate}
\end{definition}

The  \emph{decision problem} for an energy game $\mathcal{G}=\langle \Gamma,\mathcal{O}_0\rangle$ asks to solve only the first one of the two items above, i.e. to partition $V$ into $\langle W_0, W_1\rangle$.
The decision problem on energy games is equivalent to the decision problem on so called \emph{meanpayoff games}~\cite{ZP}, a game on graphs originally introduced by game theorists within the economic community, where the objective of player $0$ is to minimize the long-run average weight of plays.
Several algorithms exist to solve the decision problem on meanpayoff games (cf.~\cite{Aptbook}  for a survey of the available algorithms): indeed, it is worth noticing that the best pseudo-polynomial meanpayoff  algorithm  is based on its reduction to energy games \cite{BrimEtAl,CHN14,CR17}.  Ad-hoc procedures are instead necessary to solve  the initial credit problem on energy games, that is   specific to energy objectives.
The latter problem was solved  in \cite{BrimEtAl} with a pseudo-polynomial procedure having complexity $\mathit{\mathcal{O}(\abs{E}{\cdot}\abs{V}{\cdot}W)}$, where $\abs{E}$ (resp.~$\abs{V}$) is the number of edges in the game arena and $W$ is the maximum weight labeling an edge.
Energy games were algorithmically studied also in \cite{CHN14}, where the authors provide a polynomial algorithm for solving the initial credit problem on \EG with special weights structures.
In particular, the authors of \cite{CHN14} show that solving \EG where all the cycles are either 'good' or significantly `bad'\footnote{Graphs, for instance, where all the negative cycles have weight less than $\frac{W}{2}$, where $W$ is the maximum weight in the graph} can be done in polynomial time.

In the rest of this paper, we will show the design of a CUDA-based parallel \EG algorithm based on the procedure for the \EG initial credit problem defined  in \cite{BrimEtAl} (which is briefly described in the next subsection). The latter  allows to exploit the computational power offered by modern GPUs.

\subsection{Computing the Minimum Initial Credit of Energy  on the CPU}

In this subsection, we briefly describe the sequential algorithm in \cite{BrimEtAl} to solve the initial credit problem on energy games.
The procedure in \cite{BrimEtAl} is based on the of so called  \emph{energy progress measure},  which is recalled in Definition~\ref{defEPM} and relies on the following notation.

Let  $\preceq$ be the total order on  $\mathcal{N}=\mathbb{N}\cup \{\top\}$, where  $x \preceq y$ if and
only if either $x \leq y$ or  $y=\top$.
Let  $\ominus$ be the 
operator $\ominus: \mathcal{N} \times \zed \to \mathcal{N}$
such that, for each $a \in \mathcal{N}$ and $b \in \zed$:
$$a\ominus b=
                 \left\{
                     \begin{array}{ll}
                         \max(0,a-b) &       \mbox{if $a\neq \top$ }\\
						\top &       \mbox{otherwise}
                     \end{array}
                 \right.$$
Roughly speaking, the  \emph{local conditions} on the nodes of an energy game  $\mathcal{G}$  imposed  by an energy progress measure $f$  (cf.\ Definition~\ref{defEPM}) guarantee that the following property on $\mathcal{G}$ holds: for each node $v$ in $\mathcal{G} $, if $f(v)\neq \top$, then player $0$ has a strategy $\sigma_0$  to ensure that the energy level along each play  compatible with $\sigma_0$ is not negative, provided the initial credit $f(v)$.
\begin{definition}[Energy Progress Measure (EPM) \cite{BrimEtAl}]\label{defEPM} 
A function $f:V\to \mathcal{N}$ is a \emph{energy progress measure}
for the \EG   $\mathcal{G}=\langle \Gamma,\mathcal{O}_0\rangle$ iff the following conditions hold:
\begin{itemize}
	\item if $v\in V_0$, then $f(v)\succeq f(v')\ominus w(v,v')$ for some~$(v,v'){\in}E$
	\item if $v\in V_1$, then $f(v)\succeq f(v')\ominus w(v,v')$ for all $(v,v')\in E$
\end{itemize}
\end{definition}
For a game $\mathcal{G}$, let ${\mathcal F}$ be the set
of functions $f : V \to \mathcal{N}$, and consider the  partial order
$\sqsubseteq \subseteq {\mathcal F}\times {\mathcal F}$,  defined as $f \sqsubseteq g$
iff for all $v \in V$, $f(v) \preceq g(v)$.
The authors of \cite{BrimEtAl} proved that  $\mathcal{G}$ admits a \emph{least} energy progress measure $f$ w.r.t. $\sqsubseteq$, satisfying the following properties:
\begin{enumerate}
\item for each node $v\in W_1$, $f(v)=\top$

\item for each node $v\in V_0$, $f(v)\leq \M_{G}$, where:
$$ \M_{\mathcal{G}}=\sum_{v\in V}\max(\{0\}\cup\{-w(v,v')\mid (v,v')\in E\})$$
\end{enumerate}
Given an \EG $\mathcal{G}$, the initial energy credit algorithm in \cite{BrimEtAl} computes exactly the \emph{least energy progress measure} for $\mathcal{G}$:  $$f:V\to \mathcal{C}_\mathcal{G}=\{n\in \mathbb{N} \mid n \leq \M_{\mathcal{G}} \} \cup \{\top\}.$$
More precisely, such an algorithm  initializes $f$ to the constant function $0$ and relies on the following $\sqsubseteq$-monotone \emph{lifting  operator} to update $f$, until a least fixpoint is reached.

{\begin{algorithm}[tb]
\CommentSty{\color{blue}}
\caption{Minimum Initial Credit  Algorithm for Energy Games: Given in input an \EG $\mathcal{G}$, it computes in output the least progress measure $f: V \to \mathcal{C}_\mathcal{G}$ \label{alg:energy-games}}
 {
 \begin{flushleft}
 \Begin{
 \tcc{Initialize the set $L$ of nodes for which the least EPM  $f$ needs to be lifted}
	\nl $L \gets \{v \in V_0 \mid \forall (v,v') \in E: w(v,v') < 0 \}$ \label{alg1:line-one} \;
	\nl $L \gets L \cup \{v \in V_1 \mid \exists (v,v') \in E: w(v,v') < 0 \}$ \;
\tcc{Initialize the least EPM~$f$}

        \nl \ForEach{$v \in V$}
	{
		\nl $f(v) \gets 0$ \;
		\nl \lIf{$v \in V_0 \cap L$}{$\counter(v) \gets 0$} \;
		\nl \lIf{$v \in V_0 \setminus L$}{$\counter(v) \gets \abs{\{v' \in \post(v) \mid f(v) \geq f(v') \ominus w(v,v')\}}$} \label{alg1:before-loop} \;
	}
	\tcc{Apply the lift-operator to update the least EPM $f$ for each node in $L$}
        \nl \While{$L \neq \emptyset$ \label{alg1:while-loop}}
	{
		\nl Pick $v \in L$ \label{alg1:pick} \;
		\nl $L \gets L \setminus \{v\}$\;
\nl $old\gets f(v)$ \label{alg1:remove}\;
		\nl $f \gets \delta(f,v)$ \label{alg1:update}\;
\nl \lIf{$v \in V_0$}{$\counter(v) \gets \abs{\{v' \in \post(v) \mid f(v) \geq f(v') \ominus w(v,v')\}}$} \label{alg1:update-count} \;
        	\nl \ForEach{$v' \in \pre(v)$ {\bf such that} $f(v') < f(v) \ominus w(v',v)$ \label{alg1:for-loop}}
		{
				\nl \If{$v' \in V_0$}
				{
					\nl \lIf{$f(v') \geq old \ominus w(v',v)$} {$\counter(v') \gets \counter(v') - 1$ \label{alg1:count-update}\;}
					\nl \lIf{$\counter(v') \leq 0  $}{$L \gets L \cup \{v'\}$} \label{alg1:list-update1}\;
				}
				\nl \lIf{$v' \in V_1$}{$L \gets L \cup \{v'\}$} \label{alg1:list-update2}\;
		}
	}
	\nl \KwRet{$f$} \;
 }
\end{flushleft}
}
\end{algorithm}
}

\begin{definition}[Lifting Operator \cite{BrimEtAl}]\label{liftop}
Given $v\in V$, the lifting operator $\delta(\cdot,v):[V\to \mathcal{C}_\mathcal{G}] \to [V\to \mathcal{C}_\mathcal{G}]$ is
defined by $\delta(f,v)=g$ where:
$$g(u)=
      \left\{
          \begin{array}{ll}
              f(u) &       \mbox{if $u\neq v$}\\
              \min\{f(v')\ominus w(v,v') \mid (v,v')\in E\} &       \mbox{if $u= v\in V_0$}\\
              \max\{f(v')\ominus w(v,v') \mid (v,v')\in E\} &       \mbox{if $u= v\in V_1$}
          \end{array}
      \right.$$
\end{definition}

To conclude this subsection we report in Algorithm~\ref{alg:energy-games} the exact pseudo-code of the minimum initial credit algorithm in \cite{BrimEtAl}, based on the least energy progress measure described above. 
Such an algorithm uses suitable counters to achieve a global worst-case complexity of $\mathcal{O}(\abs{E}\cdot\M_{\mathcal{G}})$=$\mathit{\mathcal{O}(\abs{E}{\cdot}\abs{V}{\cdot}W)}$.

\section{An OpenMP implementation}\label{sec:multicore}

In order to obtain an immediate way to parallelize the computation of the  minimum initial credit on \EG, let us observe that
each application of the lift operation in Definition \ref{liftop} never decreases the value of $f(v)$ for any vertex~$v$.
Hence, processing all elements in $L$ in parallel is a sound procedure.
Moreover, as motivated in the previous subsection, a bounded number $\abs{E}\cdot\M_{\mathcal{G}}$ of lift operations suffices to determine a solution.
Consequently, a simple way to parallelize the computation of the EPM consists in applying the lift operation
 in parallel for each vertex of the graph and in iterating this step until either a fixpoint or the theoretical bound on loops is reached. 
Algorithm~\ref{alg:MPenergy-games} presents the skeleton of the resulting algorithm implemented exploiting OpenMP.
In particular, the loops starting in lines~1 (performing the initialization of $f$) and 8 (performing the lift step),
respectively, are executed in parallel by distributing
the computation among the available OpenMP threads.
The while-loop in lines 5--10
iterates until an ending condition is achieved.
We experimented with this implementation by using~1, 2, 4, and 8 threads, always mapped to different CPUs (see Section~\ref{sec:esperimenti}).

{\begin{algorithm}[tb]
\CommentSty{\color{blue}}
\caption{Naive parallel version of Algorithm~\ref{alg:energy-games}\label{alg:MPenergy-games}}
 {
 \begin{flushleft}
 \Begin{
\tcc{Initialize the least EPM~$f$}

        \nl \ForEach{$v \in V$ {\rm\bf in parallel}}
	{
		\nl $f(v) \gets 0$;
	}
	\nl $loops \gets \abs{E}\cdot\M_{\mathcal{G}};\hfill\mbox{\tt /* Bound on number of lifts */}$
\\
	\nl $more \gets true$\; 
\tcc{Apply lift-operator until fixpoint or the bound on loops is reached}
        \nl \While{$more \wedge loops>0$}
	{
		\nl $f' \gets f$\;
		\nl $loops \gets loops-1$\;
        	\nl \ForEach{$v \in V$ {\rm\bf in parallel}}
		{
			\nl $f{\gets}\delta(f,v);\hfill\mbox{\tt/* Each thread lifts a $v$ */}$
\\
			\nl {$more{\gets}more{\vee}f(v)\not=f'(v)$;\hfill\mbox{\tt/* Race condition*/}}
		}
	}
	\nl \KwRet{$f$};
 }
\end{flushleft}
}
\end{algorithm}
}

\section{A CUDA-Based Solver}\label{sec:implementazione}

{\begin{algorithm}[tb]
\CommentSty{\color{blue}}
\caption{CUDA Algorithm for Energy Games\label{alg:CUDAenergy-games}}
 {
 \begin{flushleft}
 \Begin{
 \tcc{Initialize the least EPM~$f$ and the set $L$ of nodes to be lifted}
	\nl \ForEach{$v \in V$ {\rm\bf in parallel}}
	{
		\nl $f(v) \gets 0$\;
		\nl\lIf{$(v \in V_0 \wedge (\forall (v,v') \in E: w(v,v') < 0))$
		$\vee$ $(v \in V_1 \wedge (\exists (v,v') \in E: w(v,v') < 0))$}{$L \gets L\cup\{v\}$}\;
	}
	\tcc{Apply lift-operator until a fixpoint}
        \nl \While{$L\not=\emptyset$}
	{
		\nl$L'\gets\emptyset$; ~ $f'\gets f; \hfill\mbox{\tt/* done in parallel */}$\\
        	\nl\ForEach{$v \in L$ {\rm\bf in parallel}}
		{
			\nl$f{\gets}\delta(f,v); \hfill\mbox{\tt/* Each thread lifts one $v$ */}$\\
			\nl\If{$f(v)\not=f'(v)$}
			{
        			\nl \lForEach{$v' \in \pre(v)$}
				{
					 $L' \gets L' \cup \{v'\}$\;
				}
			}
		}
		\nl$L\gets L'; \hfill\mbox{\tt/* done in parallel */}$
	}
	\nl \KwRet{$f$} \;
 }
\end{flushleft}
}
\end{algorithm}
}

This section describes the main design choices made in implementing a CUDA-based parallel solution to the \EG initial credit problem.
As concerns data structures,
the adjacency matrix of  the input EG is represented in device memory, by exploiting the standard  \emph{Compressed Sparse Row} (CSR) format,
usually employed to store sparse matrices.
The progress measure $f$ to be computed is stored in an array of $|V|$ elements.

By analyzing the pseudo-code of the sequential Algorithm~\ref{alg:energy-games}, one plainly identifies tasks that could be executed in parallel.
The simplest one is the initialization of the set~$L$ of active nodes (those whose least energy progress measure  $f$ needs to be lifted)
and the initialization of the least progress measure~$f$ (lines 1-3 in Algorithm~\ref{alg:CUDAenergy-games}).
The set $L$ is represented by an array of (at most) $|V|$ elements,
A specific kernel function has been defined for the initialization of such~$L$.
In particular, a 1-to-1 mapping (\emph{vertex-parallelism} ~\cite{jia2011edge}) assigns each node to one thread.
Each thread determines whether the corresponding node has to be inserted in $L$ or not (line~3).

The core of the sequential algorithm is in lines~7--17, where elements are extracted from $L$, one at a time, and their
progress measures are lifted.
As mentioned before, processing all elements in $L$ in parallel is a sound procedure.
Therefore, a specific kernel function has been designed to compute, in parallel, the new values of $f(v)$ for each $v$ in $L$.
In doing this, all elements $v'$ in the set $\post(v)$ have to be considered. 
To better exploit the mass parallelism supported by GPUs, each node $v$  in $L$ is assigned to a set of $h=2^k$ threads of the same warp. 
Such threads process in parallel all elements in $\post(v)$ and (conjunctively) compute the value of $\delta(f,v)$.
Information between such threads is exchanged through \emph{warp-shuffle} operations, which are
enabled because all $h$ threads always belong to the same warp.
Acting in this manner reduces the number of
accesses to global and shared memories and this, in turn, speeds-up the overall computation.
Indeed, as mentioned earlier, by means of shuffle operations data are moved directly between threads' registers
instead of communicating them through global/shared memory operations.
The value $h$ can be heuristically chosen as  (a fraction of) the average degree of nodes in~$V$.

Thanks to the use of the CSR format, all members of $\post(v)$ are stored in consecutive locations of the device memory.
This optimizes the time needed by the $h$ threads for accessing the initially needed data.
The first of the $h$ threads stores the new value of $f(v)$, after the interaction between the $h$ threads is completed.

Consider now that, by Definition~\ref{liftop}, the computation of lifting operator involves  the evaluation
of either a $\min$ operation or a $\max$ operation of a set of values, depending on the player controlling
the active node.
A further optimization is applied in order to minimize thread divergence between threads of the same warp.
The set $V$ is sorted so that all nodes in $V_0$ (resp.\ $V_1$) correspond to  consecutive lines
of the adjacency matrix of the EG.
Consequently, in all warps (but at most one) all threads always execute
the same sequence of instructions.
Namely, all of them compute the $\min$ (resp.\ $\max$) operation.

Once the progress measure $f(v)$ of a node $v$ has been updated, the set of predecessors $\pre(v)$ of $v$ has to be considered
in order to compute the new set of active nodes.
Also this task is performed in parallel by splitting the work load among the $h$ threads.
The set of $h$ threads that computed $\delta(f,v)$, process each node in $\pre(v)$
and determine if it has to be inserted in~$L$.
Notice that in this phase of the computation it might be the case that the same element might be inserted in the new $L$ because of
different reasons, as it is predecessor of different processed active nodes. 
Repeated insertion of the same element in $L$ is avoided by marking each inserted node
(a suitable vector of flags is used for this purpose).

Similarly to what done in computing $\delta(f,v)$, in order to optimize the access patterns exploited to retrieve the needed data, 
the elements of $\pre(v)$ are stored in consecutive memory locations.
This is achieved by adopting a redundant representation of~$E$.
More specifically, the  adjacency matrix of EG is
represented in the device memory using the \emph{Compressed Sparse Column} (CSC) format too.
This representation is easily computed by transposing  the corresponding CSR representation,
through standard functions provided by the CUSPARSE library.

With the differences described so far, the overall structure of the resulting CUDA implementation essentially reflects the one of the
sequential Algorithm~\ref{alg:energy-games}.
The computation starts on the CPU by reading and parsing a text file specifying the input arena.
The EG is then transferred to the device memory and a conversion from CSR to CSC is executed by the device.
Now, the CPU controls the computation by calling the device functions described earlier.
First the initialization of data is performed.
Then, the device function which improves the progress measure is repeatedly called until an empty set of active nodes
is obtained (this corresponds to the while-loop in Algorithm~\ref{alg:energy-games}). 
We experimented with different choices for the values of $h$ (the case $h=1$ clearly corresponds to vertex-parallelism, while
for $h>1$ we have warp-centric parallelism).
Finally, the result is transferred back to the host memory and output.

\begin{figure}
{\centering
        \begin{subfigure}[b]{0.70\textwidth}
                \includegraphics[width=\linewidth]{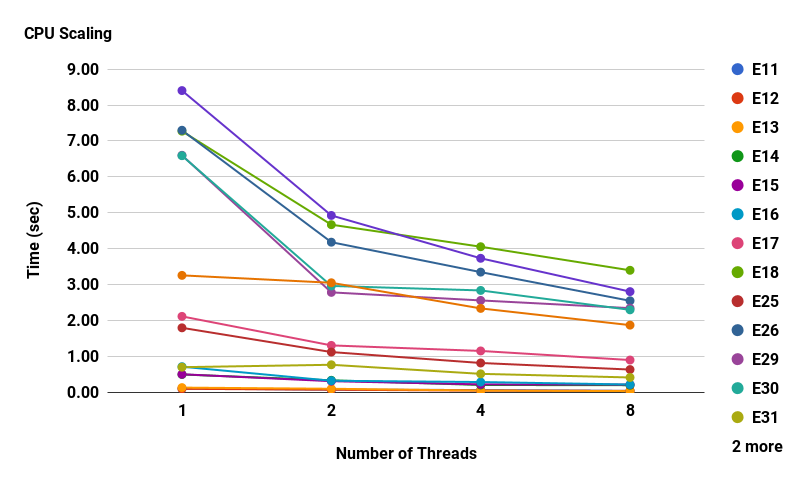}
                \caption{Dataset: Equivalence Checking.}
                \label{fig:cpuscaling-a}
        \end{subfigure}%
\\
        \begin{subfigure}[b]{0.70\textwidth}
                \includegraphics[width=\linewidth]{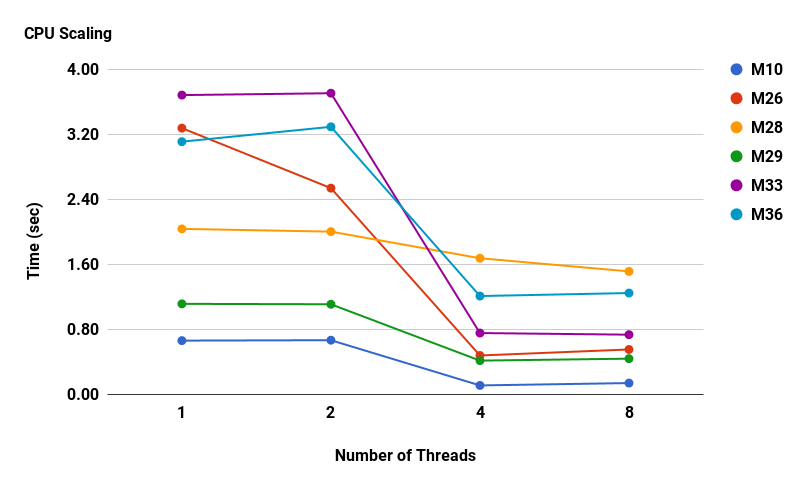}
                \caption{Dataset: Model Checking.}
                \label{fig:cpuscaling-b}
        \end{subfigure}%
\caption{CPU EG scaling experiment. The figures report the performance in seconds (y-axis) required to solve Energy Games instances by increasing the number of threads on Intel Xeon E5-2640 (x-axis).}\label{fig:cpuscaling}
}
\end{figure}

\begin{figure}{}
{\centering
                 \includegraphics[width=0.9\linewidth]{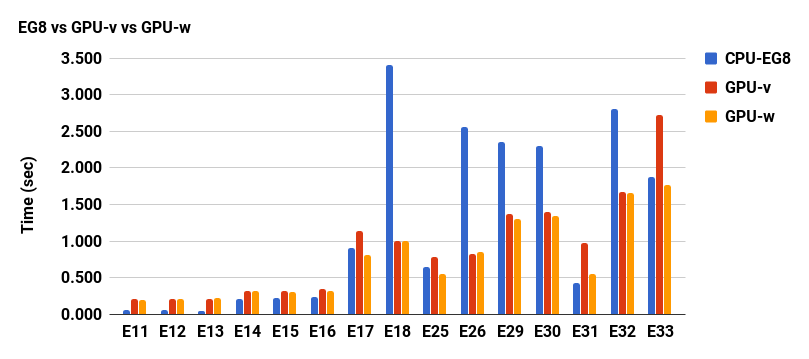}
                 \caption{Performance comparison among three different parallel implementations CPU-EG8, GPU-v and GPU-w on Equivalence Checking data set. The bars show the time-to-solution in seconds of each implementation.}
                 \label{fig:cpugpu-e}
}
\end{figure}

\begin{figure*}[]
{\centering
        \begin{subfigure}[b]{0.70\textwidth}
                \includegraphics[width=\linewidth]{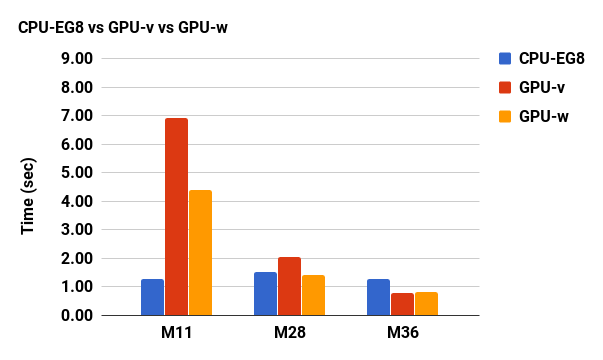}
                 \label{fig:cpugpu-m-a}
        \end{subfigure}%
\\
        \begin{subfigure}[b]{0.70\textwidth}
                \includegraphics[width=\linewidth]{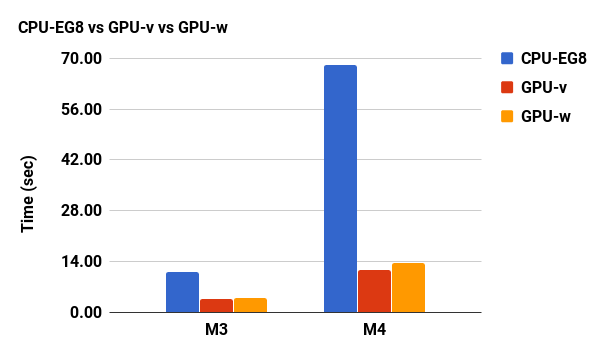}
                 \label{fig:cpugpu-m-b}
        \end{subfigure}%
\\
        \begin{subfigure}[b]{0.70\textwidth}
                \includegraphics[width=\linewidth]{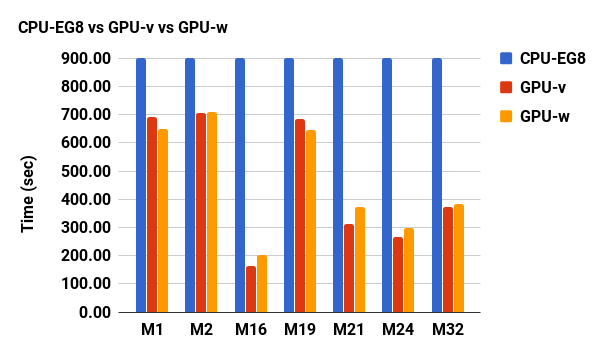}
                 \label{fig:cpugpu-m-c}
        \end{subfigure}
        \caption{Performance comparison among three different parallel implementations CPU-EG8, GPU-v and GPU-w on Equivalence Checking data set. The bars show the time-to-solution in seconds of each implementation.}\label{fig:cpugpu-m}
}
\end{figure*}

\section{Experimental Results}\label{sec:esperimenti}
Numerical experiments have been performed on a server equipped with an Intel Xeon E5-2640 v3 and four Nvidia K80 GPU. 
The code has been generated using the GNU C compiler version 4.8.2, CUDA C compiler version 7.5.

A sequential solver, named ``CPU EG1'', following the pseudo-code listed in Algorithm~\ref{alg:energy-games} has been implemented in~C. In order to develop a fair comparison with the GPU-based solution, 
the very same representation (using both CSR and CSC formats) used in the CUDA implementation has been adopted in the sequential solver. We refer to ``CPU EG2'', ``CPU EG4'' and ``CPU EG8'' the codes that implement the algorithm described in Algorithm~\ref{alg:MPenergy-games} with 2,4,8 threads respectively. 

The codename ``GPU-v'' and ``GPU-w'' denote the code implemented for GPUs based on vertex parallelism and warp parallelism respectively.
As a source for our benchmark, we consider the suite for games arena in \cite{Keiren15}. In particular, \cite{Keiren15} provides a large database of games (over $1000$ instances) that originate from different verification problems and are notable---for experimental purposes---in terms of their diversity and applicative coverability. Table~\ref{tab:summary} provides references to the filenames of the exact instances used in the experimentation as well as a succinct description of the characteristics of the graphs. Such instances encode equivalence checking ($E0$-$E33$) model-checking ($M0$-$36$) problems into qualitative games with parity objectives  \cite{Aptbook}. Standard conversion from qualitative games to quantitative games with meanpayoff and energy objectives \cite{BrimEtAl} have been used in  \cite{ML16} to generate the final data set.

\subsubsection*{Performance Analysis} 
In the present section, we show our experimental results. First, we compared the performance of CPU-EG over the data set by increasing the number of threads (strong scaling). 
Due to slow convergence time, CPU-EG is not able to solve some instances within a given time-out (for our convenience we set up it to 900 seconds). In Figure \ref{fig:cpuscaling} we only show the most representative results. Please refer to Table \ref{tab:summary} for a detailed analysis. In general, experiments show a good scalability between 2 and 4 threads, after that threads do not have enough work to do. 
For some instances (i.e., M33), CPU-EG shows a better scalability since it takes advantage from the parallelism that a more complex structure exhibits. Other instances, like M28, on the contrary, do not have a significant benefit from multi-core architectures. 

In the second set of experiments, we compare the performance (time-to-solution) between GPU-v and GPU-w. Furthermore, we also show the time of CPU-EG8 as a baseline. In detail, Figure~\ref{fig:cpugpu-e} shows the performance over ``equivalence checking'' data set, whereas Figure~\ref{fig:cpugpu-m} is related to ``model checking'' instances. Generally, CPU-EG8 is faster on ``easy'' instances where the algorithm converges quickly in few iterations. The identification of ``easy'' instances is hard to do \emph{a priori} since, as we mentioned, the convergence strongly depends on the weights and the structure of the graphs. By analyzing final performance we can say that EG-GPUs are up to $5$x than CPU-EG8 (36x faster than EG-CPU1). 
Concerning the comparison between GPU-v and GPU-w, we do not observe a significant difference in terms of performance except for a small number of instance. On average, GPU-w achieves better performance slightly up to a factor of $1.7$x. 

As a final comment, the results of this initial experimentation seem to witness the advantages offered even by a plain parallelization of Algorithm~\ref{alg:energy-games}. Although our results are remarkable, a deeper investigation has to be conducted in order to identify (if any) those classes of EG where a specific approach may achieve the best performance. Again, it seems reasonable that particular topologies of the underlying graph game may reduce the gap between the sequential and the parallel algorithms. On the other hand, various optimizations and refinements can be introduced in the parallel solver. Among the ones under consideration we just mention here the possibility of partitioning the given arena w.r.t.\ the strongly connected components of the graph game and processing them in parallel, just by imposing a topological order among the components.

\begin{table*}[bt]
{\centerline{\tiny
\begin{tabular}{|@{\hspace*{.6ex}}l@{\hspace*{.6ex}}|@{\hspace*{.6ex}}l@{\hspace*{.6ex}}|@{\hspace*{.6ex}}r@{\hspace*{.6ex}}|@{\hspace*{.6ex}}r@{\hspace*{.6ex}}|@{\hspace*{0ex}}r@{\hspace*{0.2ex}}|@{\hspace*{.6ex}}r@{\hspace*{.6ex}}|@{\hspace*{.6ex}}r@{\hspace*{.6ex}}|@{\hspace*{.6ex}}r@{\hspace*{.6ex}}|@{\hspace*{.6ex}}r@{\hspace*{.6ex}}|@{\hspace*{.6ex}}r@{\hspace*{.6ex}}|@{\hspace*{.6ex}}r@{\hspace*{.6ex}}|}
\hline
\textbf{Graph} &\textbf{ID} & \textbf{Nodes}  & \textbf{Edges}     &\textbf{\begin{tabular}{c}Avg\\Degree\end{tabular}}& \textbf{\begin{tabular}{c}CPU\\EG1\end{tabular}} & \textbf{\begin{tabular}{c}CPU\\EG2\end{tabular}}& \textbf{\begin{tabular}{c}CPU\\EG4\end{tabular}}& \textbf{\begin{tabular}{c}CPU\\EG8\end{tabular}}& \textbf{GPU-v}   & \textbf{GPU-w} \\
\hline
ABP(BW)\_Onebit\_(datasize=2\_capacity=1\_windowsize=1)eq=branching-bisim.sol               & E0       & 12239050 & 33790242  & 2.76       & 900.000 & 900.000 & 900.000 & 900.000 & 0.673   & 0.547   \\
ABP(BW)\_Onebit\_(datasize=2\_capacity=1\_windowsize=1)eq=branching-sim.sol                 & E1       & 12239050 & 33790242  & 2.76       & 900.000 & 900.000 & 900.000 & 900.000 & 0.557   & 0.554   \\
ABP(BW)\_Onebit\_(datasize=2\_capacity=1\_windowsize=1)eq=weak-bisim.sol                    & E2       & 10685466 & 40545186  & 3.79       & 900.000 & 900.000 & 900.000 & 900.000 & 0.617   & 0.558   \\
ABP(BW)\_Onebit\_(datasize=3\_capacity=1\_windowsize=1)eq=branching-bisim.sol               & E3       & 40556396 & 112380248 & 2.77       & 900.000 & 900.000 & 900.000 & 900.000 & 1.384   & 1.480   \\
ABP(BW)\_Onebit\_(datasize=3\_capacity=1\_windowsize=1)eq=branching-sim.sol                 & E4       & 40556396 & 112380248 & 2.77       & 900.000 & 900.000 & 900.000 & 900.000 & 1.331   & 1.481   \\
ABP(BW)\_Onebit\_(datasize=3\_capacity=1\_windowsize=1)eq=weak-bisim.sol                    & E5       & 35431922 & 134886692 & 3.81       & 900.000 & 900.000 & 900.000 & 900.000 & 1.542   & 1.670   \\
ABP\_Onebit\_(datasize=2\_capacity=1\_windowsize=1)eq=branching-bisim.sol                   & E6       & 9488018  & 26181506  & 2.76       & 900.000 & 900.000 & 900.000 & 900.000 & 0.473   & 0.396   \\
ABP\_Onebit\_(datasize=2\_capacity=1\_windowsize=1)eq=branching-sim.sol                     & E7       & 9488018  & 26181506  & 2.76       & 900.000 & 900.000 & 900.000 & 900.000 & 0.461   & 0.396   \\
ABP\_Onebit\_(datasize=3\_capacity=1\_windowsize=1)eq=branching-bisim.sol                   & E8       & 31611530 & 87556070  & 2.77       & 900.000 & 900.000 & 900.000 & 900.000 & 1.073   & 1.078   \\
ABP\_Onebit\_(datasize=3\_capacity=1\_windowsize=1)eq=branching-sim.sol                     & E9       & 31611530 & 87556070  & 2.77       & 900.000 & 900.000 & 900.000 & 900.000 & 1.095   & 1.075   \\
ABP\_Onebit\_(datasize=3\_capacity=1\_windowsize=1)eq=weak-bisim.sol                        & E10      & 27799634 & 104694734 & 3.77       & 900.000 & 900.000 & 900.000 & 900.000 & 1.235   & 1.307   \\
Buffer\_Onebit\_(datasize=2\_capacity=2\_windowsize=1)eq=branching-bisim.sol                & E11      & 912641   & 2762529   & 3.03       & 0.116   & 0.087   & 0.074   & 0.050   & 0.204   & 0.193   \\
Buffer\_Onebit\_(datasize=2\_capacity=2\_windowsize=1)eq=branching-sim.sol                  & E12      & 912641   & 2762529   & 3.03       & 0.109   & 0.083   & 0.058   & 0.051   & 0.205   & 0.202   \\
Buffer\_Onebit\_(datasize=2\_capacity=2\_windowsize=1)eq=weak-bisim.sol                     & E13      & 966897   & 3278913   & 3.39       & 0.142   & 0.110   & 0.060   & 0.045   & 0.209   & 0.221   \\
Buffer\_Onebit\_(datasize=3\_capacity=2\_windowsize=1)eq=branching-bisim.sol                & E14      & 3471553  & 11083645  & 3.19       & 0.507   & 0.345   & 0.215   & 0.209   & 0.311   & 0.310   \\
Buffer\_Onebit\_(datasize=3\_capacity=2\_windowsize=1)eq=branching-sim.sol                  & E15      & 3471553  & 11083645  & 3.19       & 0.513   & 0.321   & 0.231   & 0.226   & 0.310   & 0.296   \\
Buffer\_Onebit\_(datasize=3\_capacity=2\_windowsize=1)eq=weak-bisim.sol                     & E16      & 3644173  & 12968893  & 3.56       & 0.724   & 0.333   & 0.297   & 0.228   & 0.338   & 0.311   \\
CABP\_Onebit\_(datasize=2\_capacity=1\_windowsize=1)eq=strong-bisim.sol                     & E17      & 7626354  & 30467442  & 4.00       & 2.123   & 1.316   & 1.162   & 0.909   & 1.133   & 0.804   \\
CABP\_Onebit\_(datasize=3\_capacity=1\_windowsize=1)eq=strong-bisim.sol                     & E18      & 24812174 & 100409150 & 4.05       & 7.275   & 4.674   & 4.061   & 3.405   & 0.996   & 0.997   \\
Hesselink\_(Implementation)\_Hesselink\_(Specification)\_(datasize=2)eq=branching-bisim.sol & E19      & 33702306 & 76550466  & 2.27       & 900.000 & 900.000 & 900.000 & 900.000 & 0.996   & 0.996   \\
Hesselink\_(Implementation)\_Hesselink\_(Specification)\_(datasize=2)eq=branching-sim.sol   & E20      & 33702306 & 76550466  & 2.27       & 900.000 & 900.000 & 900.000 & 900.000 & 1.038   & 1.011   \\
Hesselink\_(Implementation)\_Hesselink\_(Specification)\_(datasize=2)eq=weak-bisim.sol      & E21      & 29868274 & 78747250  & 2.64       & 900.000 & 900.000 & 900.000 & 900.000 & 1.023   & 0.939   \\
Hesselink\_(Specification)\_Hesselink\_(Implementation)\_(datasize=2)eq=branching-bisim.sol & E22      & 33702306 & 76550466  & 2.27       & 900.000 & 900.000 & 900.000 & 900.000 & 0.983   & 1.033   \\
Hesselink\_(Specification)\_Hesselink\_(Implementation)\_(datasize=2)eq=branching-sim.sol   & E23      & 33702306 & 76550466  & 2.27       & 900.000 & 900.000 & 900.000 & 900.000 & 0.995   & 1.006   \\
Hesselink\_(Specification)\_Hesselink\_(Implementation)\_(datasize=2)eq=weak-bisim.sol      & E24      & 29868274 & 78747250  & 2.64       & 900.000 & 900.000 & 900.000 & 900.000 & 1.002   & 0.922   \\
Onebit\_SWP\_(datasize=2\_capacity=1\_windowsize=1)eq=strong-bisim.sol                      & E25      & 5322498  & 21604226  & 4.06       & 1.805   & 1.131   & 0.827   & 0.644   & 0.786   & 0.543   \\
Onebit\_SWP\_(datasize=3\_capacity=1\_windowsize=1)eq=strong-bisim.sol                      & E26      & 19026506 & 78220622  & 4.11       & 7.300   & 4.187   & 3.354   & 2.557   & 0.825   & 0.849   \\
Par\_Onebit\_(datasize=2\_capacity=1\_windowsize=1)eq=branching-bisim.sol                   & E27      & 10927074 & 30319218  & 2.78       & 900.000 & 900.000 & 900.000 & 900.000 & 0.514   & 0.427   \\
Par\_Onebit\_(datasize=2\_capacity=1\_windowsize=1)eq=branching-sim.sol                     & E28      & 10927074 & 30319218  & 2.78       & 900.000 & 900.000 & 900.000 & 900.000 & 0.526   & 0.426   \\
SWP\_SWP\_(datasize=2\_capacity=1\_windowsize=2)eq=branching-bisim.sol                      & E29      & 37636481 & 120755185 & 3.21       & 6.604   & 2.791   & 2.567   & 2.352   & 1.372   & 1.305   \\
SWP\_SWP\_(datasize=2\_capacity=1\_windowsize=2)eq=branching-sim.sol                        & E30      & 37636481 & 120755185 & 3.21       & 6.593   & 2.970   & 2.845   & 2.305   & 1.402   & 1.342   \\
SWP\_SWP\_(datasize=2\_capacity=1\_windowsize=2)eq=strong-bisim.sol                         & E31      & 3782172  & 11533061  & 3.05       & 0.710   & 0.777   & 0.525   & 0.424   & 0.966   & 0.551   \\
SWP\_SWP\_(datasize=2\_capacity=1\_windowsize=2)eq=weak-bisim.sol                           & E32      & 32926785 & 167527601 & 5.09       & 8.404   & 4.931   & 3.739   & 2.812   & 1.669   & 1.661   \\
SWP\_SWP\_(datasize=3\_capacity=1\_windowsize=2)eq=strong-bisim.sol                         & E33      & 14808231 & 45377590  & 3.06       & 3.264   & 3.058   & 2.345   & 1.880   & 2.720   & 1.773   \\
BRPdatasize=2\_counting.sol                                                                 & M0       & 2177202  & 2532590   & 1.16       & 0.758   & 0.751   & 0.191   & 0.213   & 0.274   & 0.273   \\
Clobberwidth=4\_height=4\_black\_has\_winning\_strategy.sol                                 & M1       & 564914   & 2185853   & 3.87       & 900.000 & 900.000 & 900.000 & 900.000 & 690.284 & 648.927 \\
Clobberwidth=4\_height=4\_white\_has\_winning\_strategy.sol                                 & M2       & 564914   & 2185853   & 3.87       & 900.000 & 900.000 & 900.000 & 900.000 & 705.454 & 708.674 \\
Hanoindisks=12\_eventually\_done.sol                                                        & M3       & 531443   & 1594321   & 3.00       & 66.898  & 66.567  & 12.706  & 11.051  & 3.510   & 3.928   \\
Hanoindisks=13\_eventually\_done.sol                                                        & M4       & 1594325  & 4782967   & 3.00       & 489.584 & 482.419 & 68.811  & 68.164  & 11.757  & 13.450  \\
Hesselinkdatasize=2\_nodeadlock.sol                                                         & M5       & 540737   & 1115713   & 2.06       & 0.010   & 0.009   & 0.002   & 0.002   & 0.179   & 0.175   \\
Hesselinkdatasize=2\_property1.sol                                                          & M6       & 1081474  & 2231426   & 2.06       & 0.071   & 0.072   & 0.057   & 0.059   & 0.218   & 0.211   \\
Hesselinkdatasize2\_property1.sol                                                           & M7       & 1081474  & 2231426   & 2.06       & 0.072   & 0.071   & 0.070   & 0.060   & 0.221   & 0.208   \\
Hesselinkdatasize=2\_property2.sol                                                          & M8       & 1093761  & 2246401   & 2.05       & 0.020   & 0.021   & 0.003   & 0.003   & 0.195   & 0.184   \\
Hesselinkdatasize=3\_nodeadlock.sol                                                         & M9       & 13834801 & 29028241  & 2.10       & 0.353   & 0.355   & 0.058   & 0.058   & 0.455   & 0.395   \\
Hesselinkdatasize=3\_property2.sol                                                          & M10      & 27876961 & 58309201  & 2.09       & 0.664   & 0.669   & 0.113   & 0.142   & 0.711   & 0.713   \\
IEEE1394nparties=2\_datasize=2\_headersize=2\_acksize=2\_property4.sol                      & M11      & 571378   & 996970    & 1.75       & 0.433   & 0.423   & 0.820   & 1.261   & 6.926   & 4.373   \\
IEEE1394nparties=2\_datasize=2\_headersize=2\_acksize=2\_property5.sol                      & M12      & 1411274  & 2454775   & 1.74       & 0.022   & 0.021   & 0.005   & 0.003   & 0.191   & 0.187   \\
Lift\_(Incorrect)nlifts=4\_nodeadlock.sol                                                   & M13      & 998790   & 5412890   & 5.42       & 0.192   & 0.189   & 0.153   & 0.165   & 0.237   & 0.241   \\
Lift\_(Incorrect)nlifts=4\_safety\_1.sol                                                    & M14      & 788879   & 4146139   & 5.26       & 0.051   & 0.051   & 0.006   & 0.005   & 0.199   & 0.200   \\
Onebitdatasize=3\_invariantly\_infinitely\_many\_reachable\_taus.sol                        & M15      & 867889   & 4933009   & 5.68       & 0.096   & 0.096   & 0.020   & 0.019   & 0.208   & 0.207   \\
Onebitdatasize=3\_messages\_read\_are\_inevitably\_sent.sol                                 & M16      & 579745   & 3354841   & 5.79       & 900.001 & 900.000 & 900.000 & 900.000 & 162.310 & 201.437 \\
Onebitdatasize=3\_no\_duplication\_of\_messages.sol                                         & M17      & 1191962  & 6907934   & 5.80       & 0.126   & 0.124   & 0.163   & 0.158   & 0.253   & 0.261   \\
Onebitdatasize=3\_no\_spontaneous\_messages.sol                                             & M18      & 1278433  & 7843609   & 6.14       & 0.080   & 0.080   & 0.008   & 0.009   & 0.234   & 0.230   \\
Onebitdatasize=3\_read\_then\_eventually\_send.sol                                          & M19      & 1350433  & 7068673   & 5.23       & 900.000 & 900.000 & 900.000 & 900.000 & 685.262 & 644.088 \\
SWPdatasize=2\_windowsize=3\_infinitely\_often\_receive\_for\_all\_d.sol                    & M20      & 588868   & 2071109   & 3.52       & 0.189   & 0.196   & 0.039   & 0.039   & 0.198   & 0.197   \\
SWPdatasize=2\_windowsize=3\_invariantly\_infinitely\_many\_reachable\_taus.sol             & M21      & 670177   & 2375809   & 3.55       & 900.000 & 900.000 & 900.000 & 900.000 & 311.157 & 373.967 \\
SWPdatasize=2\_windowsize=3\_no\_duplication\_of\_messages.sol                              & M22      & 944090   & 3685946   & 3.90       & 0.124   & 0.126   & 0.069   & 0.071   & 0.366   & 0.269   \\
SWPdatasize=2\_windowsize=3\_read\_then\_eventually\_send\_if\_fair.sol                     & M23      & 586658   & 1782434   & 3.04       & 0.059   & 0.060   & 0.017   & 0.018   & 0.191   & 0.200   \\
SWPdatasize=2\_windowsize=3\_read\_then\_eventually\_send.sol                               & M24      & 917713   & 3283153   & 3.58       & 900.000 & 900.000 & 900.000 & 900.000 & 265.119 & 298.995 \\
SWPdatasize=2\_windowsize=4\_infinitely\_often\_receive\_d1.sol                             & M25      & 3487362  & 12463874  & 3.57       & 1.211   & 1.207   & 0.232   & 0.235   & 0.329   & 0.331   \\
SWPdatasize=2\_windowsize=4\_infinitely\_often\_receive\_for\_all\_d.sol                    & M26      & 6974724  & 24927749  & 3.57       & 3.277   & 2.539   & 0.481   & 0.555   & 0.478   & 0.482   \\
SWPdatasize=2\_windowsize=4\_nodeadlock.sol                                                 & M27      & 2589057  & 11565569  & 4.47       & 0.095   & 0.094   & 0.015   & 0.014   & 0.267   & 0.262   \\
SWPdatasize=2\_windowsize=4\_no\_duplication\_of\_messages.sol                              & M28      & 11488274 & 45840722  & 3.99       & 2.038   & 2.004   & 1.677   & 1.515   & 2.047   & 1.388   \\
SWPdatasize=2\_windowsize=4\_read\_then\_eventually\_send\_if\_fair.sol                     & M29      & 7310722  & 22667778  & 3.10       & 1.116   & 1.111   & 0.418   & 0.442   & 0.428   & 0.433   \\
SWPdatasize=4\_windowsize=2\_infinitely\_often\_receive\_for\_all\_d.sol                    & M30      & 653574   & 2444297   & 3.74       & 0.220   & 0.218   & 0.045   & 0.047   & 0.204   & 0.200   \\
SWPdatasize=4\_windowsize=2\_no\_duplication\_of\_messages.sol                              & M31      & 858114   & 3433490   & 4.00       & 0.062   & 0.062   & 0.058   & 0.057   & 0.281   & 0.235   \\
SWPdatasize=4\_windowsize=2\_read\_then\_eventually\_send.sol                               & M32      & 869569   & 3200129   & 3.68       & 900.000 & 900.000 & 900.000 & 900.000 & 371.087 & 381.734 \\
SWPdatasize=4\_windowsize=3\_infinitely\_often\_receive\_d1.sol                             & M33      & 8835074  & 34391042  & 3.89       & 3.683   & 3.706   & 0.757   & 0.737   & 0.618   & 0.658   \\
SWPdatasize=4\_windowsize=3\_nodeadlock.sol                                                 & M34      & 7429633  & 32985601  & 4.44       & 0.309   & 0.313   & 0.030   & 0.043   & 0.445   & 0.446   \\
SWPdatasize=4\_windowsize=3\_no\_generation\_of\_messages.sol                               & M35      & 6690605  & 29413398  & 4.40       & 0.292   & 0.292   & 0.046   & 0.039   & 0.432   & 0.429   \\
SWPdatasize=4\_windowsize=3\_read\_then\_eventually\_send\_if\_fair.sol                     & M36      & 24565250 & 73675010  & 3.00       & 3.110   & 3.293   & 1.212   & 1.249   & 0.779   & 0.790  \\
\hline
\end{tabular}
}\caption{Dataset characteristics and experimental results. The first four columns show the instance name, a short ID used in the paper to refer to them, as well as some characteristics of the graphs such as size and vertex degrees. 
The rightmost six columns report the timings obtained by the different implementations described in the paper. In particular,  the four columns labeled ``CPU EG$n$'', list the results obtained by the sequential Algorithm~\ref{alg:energy-games} implemented in~C (``CPU EG1''), and by its OpenMP version (Algorithm~\ref{alg:MPenergy-games}) for different number of threads (namely, for $n=2,4,8$). The columns ``GPU-v'' and ``GPU-w'' show the results obtained by the CUDA based solver (cf., Algorithm~\ref{alg:CUDAenergy-games}), exploiting a pure vertex parallelism and warp parallelism, respectively. All timings are in seconds. A timeout of 15 minutes has been applied (the values $900.000$ denote timeout expiration).}\label{tab:summary}}
\end{table*}

\section{Related Works}\label{related}
Similar approaches exist for other kind of games used in the context of  computer aided design and formal verification. For instance, the parallelization of Meanpayoff Games has been dealt with
in~\cite{Chaloupka} and in~\cite{HoffmannL13}. Whereas in the first case the target architecture is not
GPU-based, but it is a common multi-core machine, \cite{HoffmannL13} proposes an OpenCL implementation suitable to run on AMD devices.
A proposal concerning Parity Games has been described in~\cite{ML16}, also based on OpenCL.

Several solutions have been proposed to reduce the workload unbalancing among threads and alleviate the irregular memory access. 
Jia {\rm et al.} \cite{jia2011edge} evaluated two different data-thread mapping techniques \textit{vertex-parallel} and \textit{edge-parallel}.
Due to the difference in the out-degree among vertices in scale-free networks, \textit{vertex-parallel} suffers from load imbalance among threads.
The \textit{edge-parallel} approach solves that problem by assigning edges to threads during the frontier expansion.
However, it is not suitable for graphs with a low average degree, as well as dense graphs \cite{jia2011edge}.
Furthermore, the edge-based parallelism requires much memory and atomic operations \cite{jia2011edge,suriyuce2015} especially for Energy Games instances where an {\emph{atomic min}} can be required. 
Mclaughlin and Bader~\cite{mclaughlin2014scalable} discussed two hybrid methods for the selection of the parallelization strategy.
Sarıy\"{u}ce et al.~\cite{suriyuce2013}, introduced the vertex virtualization technique based on a relabeling of the data structure (e.g., CSR, Compressed Sparse Row). 
The technique replaces a high-degree vertex $\textit{v}$ with $n_v = \lceil \texttt{adj(v)} \rceil / \Delta $ virtual vertices
having at most $\Delta$ neighbors.
Vertex virtualization technique is not very effective for graphs with a low average degree.
Typical Energy Games instances are characterized by a low average out degree (cfr. Table ~\ref{tab:summary}),  therefore a vertex-based parallelism would be more suitable for such instances.
Other efficient data-thread mapping techniques, like {\emph{active-edge parallelism}}~\cite{BERNASCHI2015145, Bernaschi2016} or other warp-centric strategies \cite{Hong:2011:ACG:1941553.1941590}, seem to be not very effective for Energy games instances where the average degree is pretty low. 

\section{Concluding Remarks}\label{conclusions}
To the best of our knowledge, we present the first GPU-based implementation of a solver for Energy Games. 
We investigated the possibility of implementing a solver for the initial credit problem on Energy Games capable of exploiting the computational power offered by modern Graphics Processing Units. 
We illustrated how a first prototype relying on the SIMT conceptual model of parallelism adopted  within CUDA framework can be plainly obtained by parallelizing the different steps of a sequential algorithm.
The proposed CUDA-based solver tersely exhibits great performance and demonstrated the viability of the approach, when compared against its sequential and CPU multi-core counterpart.
However, a detailed analysis of the topology of the graph is still required in order to design an efficient data-thread mapping technique on GPUs. 
Further, a number of improvements and heuristics can be applied to our current implementation, involving for example a static analysis of the input instance aimed at customizing the configuration parameters used to launch the CUDA kernels, or aimed at taking advantage of the topological structure of the graph. These are challenging themes for future work.


\end{document}